\documentclass[12pt,preprint]{aastex}

\input psfig.sty

\begin{document}

\title{Ultra High Energy Cosmic Rays: Strangelets?$^{0\dag}$\\
----- Extra dimensions, TeV-scale black holes and strange matter}

\author{XU Renxin, WU Fei\\
        School of Physics, Peking University,
        Beijing 100871, China}

\altaffiltext{0}{$^{\dag}$This work is supported by NSFC (No.
10273001), by the Special Funds for Major State Basic Research
Projects of China (G2000077602), and by Chinese Undergraduate
Research Endowment in Peking University.}

\begin{abstract}

The conjecture that ultra high energy cosmic rays (UHECRs) are
actually strangelets is discussed.
Besides the reason that strangelets can do as cosmic rays beyond
the GZK-cutoff, another argument to support the conjecture is
addressed in this letter via the study of formation of TeV-scale
microscopic black holes when UHECRs bombarding bare strange stars.
It is proposed that the exotic quark surface of a bare strange
star could be an effective astro-laboratory in the investigations
of the extra dimensions and of the detection of ultra-high energy
neutrino fluxes.
The flux of neutrinos (and other point-like particles) with energy
$>2.3\times 10^{20}$ eV could be expected to be smaller than
$10^{-26}$ cm$^{-2}$ s$^{-1}$ if there are two extra spatial
dimensions.

\vspace{0.4cm} %
\noindent %

\end{abstract}

\section*{}

Two puzzles in modern physics and astrophysics at least: What's
the nature of ultra high energy cosmic rays (UHECRs)? Does strange
(quark) matter exist?
These two might be understood by offering a simple suggestion that
UHECRs are actually strange matter, which does not conflict with
the study of extra dimensions.

The UHECRs\cite{wx02} detected, with energy as high as $3\times
10^{20}$ eV, can not be usual cosmic rays (protons, nuclei) due to
the GZK-cutoff, and also not be photons since photon-photon pair
production can significantly loss the energy, and various ideas
appear then in literatures to understand the observations,
including decaying of topological defects, violation of the
Lorentz invariance, etc.
However the more realistic candidates, within the framework of the
standard model of particle physics, are neutrinos\cite{fs02} and
strangelets\cite{ml02}, both of which are basically unaffected by
the GZK-cutoff.
Madsen \& Larsen\cite{ml02} suggested that UHECRs are strangelets
(stable lumps of strange matter) since strangelets can have high
mass (circumventing the GZK-cutoff) and charge (being helpful for
acceleration) but low charge-to-mass ratio.
In addition, many detected events (e.g., the Centauro events) of
cosmic ray experiments were suggested to be
strangelet-originated\cite{ww96}.
But a necessary and difficult issue is the detailed Monte Carlo
simulation of strangelet shower development, which is certainly
important to obtain a certain conclusion.
Actually the propagation of strangelets through the terrestrial
atmosphere is considered\cite{bgrs99,bgrs00,bgmrs00}, by which
some exotic cosmic ray events may be explained.
If UHECRs are strangelets, they are very probably not neutrinos
since maybe no Greisen neutrinos (through the interaction of
UHECRs with the cosmic microwave background) can be produced.

It is of fundamental importance to study strange matter in physics
and astrophysics\cite{suzhou02}. Strange matter may exist if the
Bodmer-Witten's conjecture is correct, while the most essential
thing is how to find convincing evidence in lab-physics and/or
astrophysics.
Previously it is a common opinion that strange stars are crusted,
but this concept was criticized by Xu \& Qiao\cite{xq98,xian02},
who addressed that {\em bare} strange stars (BSSs, i.e., strange
stars without crusts) being chosen as the interior of pulsars have
advantages.
Up to now, there is actually possible evidence for
BSSs\cite{xian02}: drifting sub-pulses of radio pulsars,
featureless thermal spectrum of compact stars, and super-Eddington
luminosity of soft $\gamma$-ray repeaters.
Besides helping to identify strange stars, the bare quark surface
can also be valuable in the study of the formation of TeV-scale
black holes.
Such mini black hole formation is another astrophysical
consequence of BSSs, and we will try to demonstrate below that
neutrinos as the candidates of UHECRs can also be excluded at
least in the case of two extra dimensions.

TeV-scale black hole is an addition to the old black hole family
(primordial black hole, stellar black hole, and supermassive black
hole).
Recently a great deal of attention is paid to the possibility that
our space has more than three dimensions\cite{ruba01,land02},
especially after Arkani-Hamed et al.'s\cite{add98,aadd98}
suggestion that the compactified extra dimensions could be as
large as $\sim 1$ mm.
It is well known that the Plank scale is $M_{\rm pl}=\sqrt{\hbar
c/G}\simeq 1.2\times 10^{19}$GeV/$c^2$ ($\hbar$ if the Plank
constant, $c$ is the speed of light, and $G$ is the Newtonian
gravitational constant), but $M_{\rm pl}$ may meaningless if the
space of our universe is actually of $D=3+n$ dimensions, with $n$
extra spatial dimensions.
The fundamental gravity scale, $M_*$, for $D$ spacial dimensions
is then
\begin{equation}
M_*^{n+2}\simeq ({\hbar\over c})^n M_{\rm pl}^2 R^{-n}, %
\label{M*}
\end{equation}
if the $n$-dimensional extra space is flat and compact with radii
of the order of $R$.
Arkani-Hamed et al.\cite{add98} assume $M_*\sim 1$ TeV in order to
solve the hierarchy problem of the standard model, the problem of
why there exists such a large ``desert'' between the electroweak
scale (of order $M_{\rm EW}\sim 1$ TeV) and the Plank scale
($M_{\rm pl}\sim 10^{16}$ TeV).
One can obtain $R\simeq (\hbar/M_*c)(M_{\rm pl}/M_*)^{2/n}$ from
eq.(\ref{M*}). The extra dimensions should be $n\geq 2$, because
of $R=2.8\times 10^{15}$ cm for $n=1$ which conflicts with
observations.
In the string theory, the extra dimensions can be as large as
$n=7$, we thus list the compact radii in case of $M_*=1$ TeV for
indications: $R=0.24$ cm for $n=2$, $R=1.0\times 10^{-6}$ cm for
$n=3$, $R=2.2\times 10^{-9}$ cm for $n=4$, $R=5.3\times 10^{-11}$
cm for $n=5$, $R=4.5\times 10^{-12}$ cm for $n=6$, and
$R=7.8\times 10^{-13}$ cm for $n=7$.
Only black holes with mass $M>M_*$ may be expected to form since
many unknown quantum gravity effects (e.g., the string
excitations) could play important role for $M<M_*$.
The Schwartzchild radius for a spacial D-dimensional, neutral,
non-rotating black hole with mass $M_{\rm BH}$ is\cite{adm98}
\begin{equation}
r_{\rm s}={\hbar\over \sqrt{\pi}cM_*}[{M_{\rm BH}\over M_*}
{8\Gamma({n+3\over 2})\over n+2}]^{1\over n+1}. %
\label{rs}
\end{equation}
For TeV-scale black holes, $M_{\rm BH}\sim M_*\sim 1$ TeV, one has
$r_{\rm s}\sim (1.54,1.49,1.50,1.54,1.58,1.63)\times 10^{-17}$ cm
for $n=2,3,4,5,6,7$, respectively.
It is found that $R$ is much larger than $r_{\rm s}$.

If UHECRs, with energy $\ga 10^{19}$eV, are structureless
point-like particles in the standard model, miniature black holes
may form when they bombard BSSs\cite{gll02}.
Without losing generality, let's assume UHECRs are neutrinos, and
have a discussion below.
When such a neutrino with energy $E_\nu$ interacts with a quark
with mass $m_{\rm q}$ in a BSS, a TeV-scale black hole may form if
the center-of-mass energy $E_{\rm cm}=\sqrt{2c^2m_{\rm
q}E_\nu}>M_*\sim 1$ TeV and the interaction is within a scale of
$\sim r_{\rm s}$ ($M_{\rm BH}c^2\sim E_{\rm cm}$).
As the neutrino and the quark are extremely close, with a scale of
$R$, they may feel extra dimensions by much strong gravitational
interaction since gravitons can propagate in bulk although most of
the other particles are in brane only.
Once a TeV-scale mini black hole forms, it
decays\cite{tu02,gll02}, radiating thermally, over a surface area
$A$, at the Hawking temperature $T_{\rm H}$,
\begin{equation}
A=r_{\rm s}^{n+2}\cdot {2\pi^{(n+3)/2}\over \Gamma({n+3\over 2})},
~~~T_{\rm H}(M_{\rm BH})={\hbar c\over k}\cdot {n+1\over 4\pi r_{\rm s}}, %
\label{TH}
\end{equation}
where $k$ is the Boltzmann constant, unless $T_{\rm H}$ is smaller
enough\cite{gll02},
\begin{equation}
T_{\rm H}(E_{\rm cm}/c^2)<T_{\rm eff}\equiv
T_{\rm F}\sqrt{\gamma}[1+(1-\gamma^{-2})/3]^{1/4}, %
\label{Teff>TH}
\end{equation}
where $T_{\rm F}$ is the Fermi temperature of quarks, and
$\gamma=(E_\nu+m_{\rm q}c^2)/E_{\rm cm}$ is the Lorentz factor of
the newborn mini black hole after the initial collision.
The initially produced hole increases mass by absorption of
another particle if Eq.(\ref{Teff>TH}) is satisfied, and will
continue to accrete to a mass of $E_\nu$, in a scale ($\sim 0.1$
mm) being much smaller than the radius of a BSS, before it
stops\cite{gll02} (i.e., with $\gamma=1$).
At this time, the mini black hole causes eventually a catastrophic
collapse of all of the BSS into a stellar black hole if $T_{\rm
H}(E_\nu/c^2)<T_{\rm F}$.
Therefore neutrino-induced collapse of BSSs requires both $T_{\rm
H}(E_{\rm cm}/c^2)<T_{\rm eff}$ and $T_{\rm H}(E_\nu/c^2)<T_{\rm
F}$.

\begin{figure}

\centerline{\psfig{figure=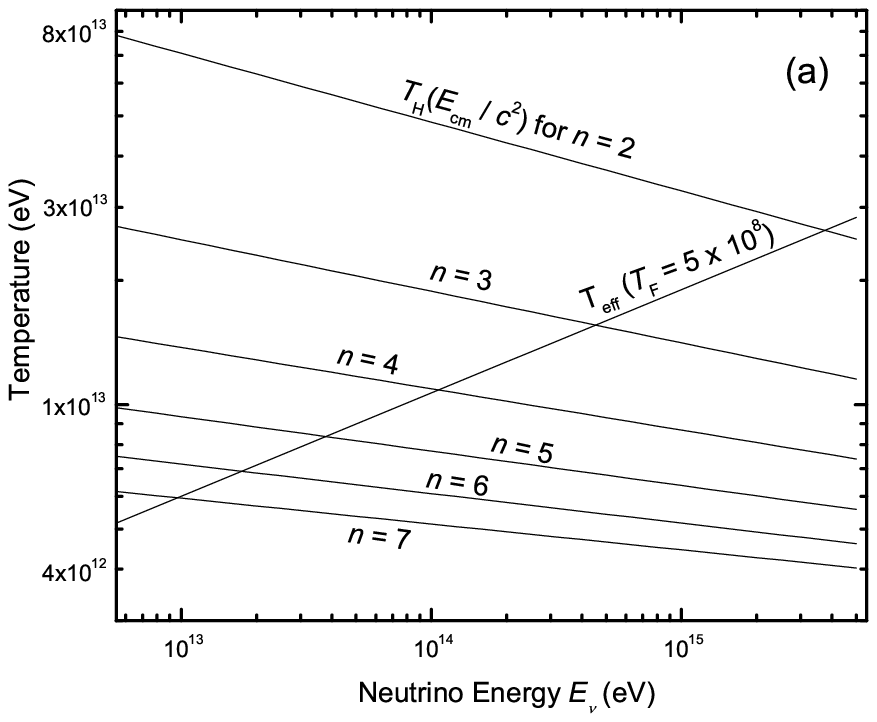,height=80mm,width=100mm,angle=0}}
\vspace{5mm}
\centerline{\psfig{figure=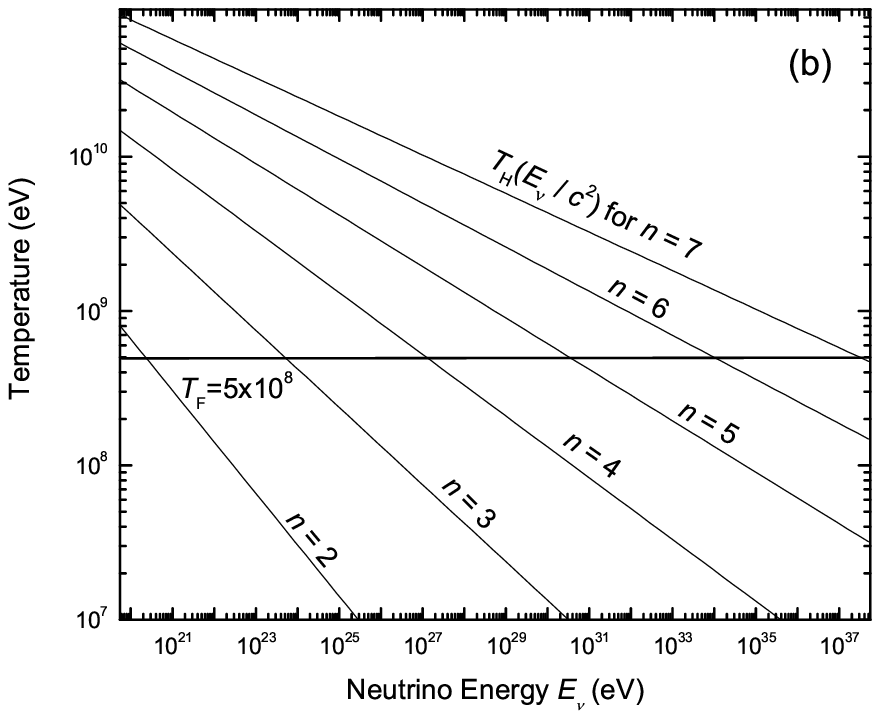,height=80mm,width=100mm,angle=0}}
\caption{
Black hole temperature ($T_{\rm H}$) and environment temperature
($T_{\rm eff}$ and $T_{\rm F}$) vs. neutrino energy ($E_\nu$).
(a). A TeV-scale black hole can form if $T_{\rm H}(E_{\rm
cm}/c^2)<T_{\rm eff}$, through an initial collision between the
neutrino and a quark.
(b). The whole BSS may collapse into a stellar black hole if
$T_{\rm H}(E_\nu/c^2)<T_{\rm F}$ by effective accretion.
In the calculation, we have $M_*=1$ TeV, $T_{\rm F}=0.5$ GeV, and
$m_{\rm q}=m_{\rm s}=200$ MeV. %
\label{Fig1}}
\end{figure}

We calculated for these two requirements by choosing $M_*=1$ TeV,
$T_{\rm F}=0.5$ GeV, and $m_{\rm q}=m_{\rm s}=200$ MeV ($m_{\rm
s}$ is the current mass of strange quark); the results are shown
in Fig.\ref{Fig1}.
It is found that, for $n$ from $2$ to $7$, the neutrino energy at
which $T_{\rm H}(E_{\rm cm}/c^2)<T_{\rm eff}$ is fulfilled is much
lower than that where $T_{\rm H}(E_\nu/c^2)<T_{\rm F}$ is
satisfied.
This means that it is easier for an ultra high energy neutrino to
trigger the formation of a TeV-scale black hole which may
eventually decay by Hawking radiation, whereas neutrino-induced
collapse of the whole star into a stellar black hole needs much
higher $E_\nu$.
So one can have following a scenario for a neutrino with
$E_\nu>(10^{13}\sim 10^{15})$ eV when it collides a BSS: it either
leads formation of a TeV-scale black hole which evaporates soon,
or results in a collapse of the BSS into a stellar black hole if
$E_\nu$ is high enough.
However for such a neutrino bombarding a neutron star or a crusted
strange star, only a TeV-scale black hole with substantial Hawking
radiation can be created and the formation of a stellar black hole
is impossible, since the mean free length ($\sim \rho r_{\rm
s}^2/m_{\rm u}\sim 1$ cm, with the outer-crust density $\rho\sim
10^{10}$g/cm$^3$ and the atomic mass unit $m_{\rm u}$) of the
black hole production is much smaller than the thickness ($\sim
0.5$ km) of the crusts\cite{gll02}.
It is worth noting that the extra dimensions are supposed to be
flat in the above calculations. A study of warped extra
dimensions, as in the model of Randal \& Sundrum, is necessary,
which may results in the suppression of black hole
growth\cite{fe02}.

The exotic quark surface of a bare strange star could thus be an
effective astro-laboratory in the investigations of the extra
dimensions and of the detection of ultra-high energy neutrino
fluxes.
Generally the age of BSSs should be greater than $10^{7\sim 9}$
year if BSSs are the nature of pulsars. This gives a limit of
neutrino flux smaller than $10^{-26}$ cm$^{-2}$ s$^{-1}$ for
\{$E_\nu>2.3\times 10^{20}$ eV, $n=2$\} (\{$E_\nu>5.1\times
10^{23}$ eV, $n=3$\}, \{$E_\nu>1.3\times 10^{27}$ eV, $n=4$\},
\{$E_\nu>3.5\times 10^{30}$ eV, $n=5$\}, \{$E_\nu>1.0\times
10^{34}$ eV, $n=6$\}, \{$E_\nu>3.2\times 10^{37}$ eV, $n=7$\}),
based on Fig.\ref{Fig1}b.
This upper limit conflicts with observation if $n=2$, and the
UHECRs events with energy $\ga 10^{20}$ eV motivated Gorham et
al.\cite{gll02} to address that $n=2$ is excluded. Nonetheless,
because of the advantages\cite{ml02} of strangelets as UHECRs (1,
higher electric charge helps to accelerate to much higher energy;
2, more massive strangelets can easily go beyond the GZK-cutoff),
one can not simply ruled out $n=2$.
If our universe has really two extra dimensions, the observed
events may in fact be a hint of UHECRs being strangelets (rather
than structureless particles, e.g., neutrinos), since increasing
possible evidence for BSSs appears\cite{xian02}.
Furthermore, if one can identify astrophysical events, with rate
of ${\cal R}_{\rm BSS\rightarrow BH}$, of neutrino-induced
collapse of BSSs to black holes in the future, we could obtain
observational constrain on the extra dimensions by combining the
studies of the neutrino spectrum (if being known) and of ${\cal
R}_{\rm BSS\rightarrow BH}$.
It is worth noting that, if the recently discovered UHECRs with
energy $\sim 10^{20}$ eV are actually strangelets, maybe no
TeV-scale black hole will be found in future neutrino
detectors\cite{hh02} (e.g., ICECUBE, RICE), at least the event
numbers would be much smaller than that expected
previously\cite{fs02}.
We note here that, Kravchenko et. al.\cite{kra02} find actually no
clear event of ultra high energy neutrino interacting with ice by
analyzing the date from the RICE antenna array, but put only upper
limits on the flux of such neutrinos, which could be significantly
smaller than the results of Fly's Eye and AGASA (see Fig.8 in
\cite{kra02}).

What is the astrophysical origin of strangelets with ultra high
energy?
One possibility of creation is during supernova exploration since
strangelets produced at a very early history of the Universe would
have evaporated a long time ago\cite{af85}. But little theoretical
work is done to understand the strangelet production in this way,
including the mass distribution of strangelets ejected from a
protostrange star and the full emerging spectrum in mass-energy
after the strangelets pass through the expanding shell\cite{vh98}.
Strong magnetic field ($\sim 10^{12}$ G) is created
soon\cite{xb01}, and a strangelet with baryon number $A\sim 10^9$
can be accelerated to $\sim 10^{20}$ eV in the unipolar induced
electric field ($\sim 6\times 10^{16}/P_{10}^2$ volts, with
$P_{10}$ the initial rotation period in 10 ms, $P_{10}\sim 1$) if
the strangelet have charge\cite{ml02} $Z\simeq 8A^{1/3}$ and is
almost totally ionized.
The Lorentz factor of a strangelet with baryon $A\sim 10^9$ and
energy $\sim 10^{20}$ eV is only $\gamma \sim 10^2$, and the
radiative energy losses proposed\cite{m03} are thus negligible
because of high baryon numbers (the radiation efficiency is
proportional to $(Z/A)^2\sim A^{-2/3}$).
The energy per quark in such a strangelet is only $\sim 10^{11}$
eV, which is too low to trigger the formation of a TeV-scale black
hole.

There are some other astrophysical indications of strange matter
ejected during supernova explorations.
For example, pulsar planets were discovered\cite{wf92} but their
astrophysical origin is not known with certainty\cite{mh01}; an
alternative and simple suggestion is that they are in fact strange
(matter) objects which were ejected with a velocity being smaller
than the escaping velocity from stellar surface and then fell back
to planet orbits.
Strange objects with mass much smaller than planet one, in a
fossil disk formed after supernova exploration, may sometimes
accrete onto the center star as accretion flow falls toward the
star. If the star is a BSS, the gravitational energy release in
this process may trigger an extremely super-Eddington burst, like
that observed in soft $\gamma$-ray repeaters (SGRs).
As addressed\cite{zxq00,usov01}, it may be natural to explain the
burst, with peak luminosity $\sim 10^7$ times of Eddington one,
and the light curves in a framework that a comet-like object falls
to a BSS.
We propose that these comet-like objects are actually strange
objects. For such an object with $\sim 10^{24}$ g, its radius is
$\sim 10^8$ cm if its density $\sim 1$ g/cm$^3$, but is only $\sim
10$ m if it is a strange object.
As we know pulsar-like stars have a typical radius $\sim 10^6$ cm,
because of the strong tide effect near the star, the comet-like
objects can not be composed of water, dust, or other ordinary
matter, but of strange matter.
In addition, the gravitational microlensing study reports objects
with much low mass\cite{sahu01}, which may be composed also of
strange matter.

\end{document}